\documentclass[aps,prb,twocolumn,showpacs,preprintnumbers,amsmath,amssymb]{revtex4}
\usepackage{graphicx}
\usepackage{hyperref}
\usepackage{natbib}
\usepackage{dcolumn}
\usepackage{bm}

\begin{document}

\title{Hydrogen on graphene: Electronic structure,
total energy, structural distortions, and magnetism
from first-principles calculations}

\author{D. W. Boukhvalov}
\email{D.Bukhvalov@science.ru.nl} \affiliation{Institute for
Molecules and Materials, Radboud University of Nijmegen, NL-6525
ED Nijmegen, the Netherlands}
\author{M. I. Katsnelson}
\affiliation{Institute of Molecules and Materials, Radboud
University of Nijmegen, NL-6525 ED Nijmegen, the Netherlands}
\author{A. I. Lichtenstein}
\affiliation{Institute of Theoretical Physics, University of Hamburg,
20355 Hamburg, Germany}

\date{\today}


%
%
\begin{abstract}
Density functional calculations of electronic structure, total
energy, structural distortions, and magnetism for hydrogenated
single-layer, bilayer, and multi-layer graphene are performed. It
is found that hydrogen-induced magnetism can survives only at very
low concentrations of hydrogen (single-atom regime) whereas
hydrogen pairs with optimized structure are usually nonmagnetic.
Chemisorption energy as a function of hydrogen concentration is
calculated, as well as energy barriers for hydrogen binding and
release. The results confirm that graphene can be perspective
material for hydrogen storage. Difference between hydrogenation of
graphene, nanotubes, and bulk graphite is discussed.
\end{abstract}
\narrowtext \maketitle
\clearpage

\section{Introduction}

Discovery of graphene, the first truly two-dimensional crystal,
and its exotic electronic properties (for review, see Refs.
\onlinecite{graphene1, graphene2,graphene3}) initiates a huge
growth of interest to carbon materials. Most of activity is
focused on electronic transport phenomena in graphene, keeping in
mind potential applications for carbon-based electronics. However,
chemical physics of graphene is also very interesting, in
particular, due to opportunity to use graphene for chemical
sensors with extraordinary sensitivity \cite{graphene4}. Another
interesting direction of investigations is a possible use of
graphene for hydrogen storage. One could expect that
two-dimensional systems could be very convenient for this aim.

In general, carbon-based systems are among the most attractive
objects for hydrogen storage \cite{Dillon2}. A promising storage
properties of single-wall carbon nanotubes (SWCNT) were first
reported in Ref. \onlinecite{Dillon1}. In last few year graphene
was used as a model system to study the electronic structure and
adsorption properties of the SWCNT \cite{Leht, Duplock}. After
discovery of real graphene several works appeared theoretically
studying the hydrogen adsorption on graphene, as a special
material (see, e.g., Refs. \onlinecite{Yazyev1, Ito-MD}). It is
commonly accepted now \cite{Leht, Duplock, Yazyev1, Ito-MD} that
the chemisorption of single hydrogen atom on graphene leads to
appearance of magnetic moments in the system. The magnetic
interactions between the hydrogen atoms placed at large distances
on graphene have been calculated in Ref. \onlinecite{Yazyev1}.
However, energetics of various hydrogen configurations taking into
account carbon sheet relaxation was not studied yet in detail. In
earlier works, only a very special structure of hydrogenated
graphene, with all hydrogens sitting on the same side was
discussed. Here we will demonstrate that actual energetically
favorable structure with hydrogenization of the both sides has
quite different properties and, in particular, turns out to be
nonmagnetic.

Earlier a similar structure has been discussed for the case of
SWCNT \cite{bi1, bi2}. However, in contrast with the SWCNT in
graphene there is no specific potential barrier for hydrogen atoms
\cite{bi2} since both sides of graphene are equally achievable for
the adsorption which makes the situation different. Deeper
understanding of the case of graphene will be useful also to
discuss hydrogen storage capacity of nanotubes \cite{Leht,
Duplock} or nonporous carbon \cite{vacan}, as well as
corresponding experimental results for graphite \cite{magnogr}.
Effect of curvature on the hydrogen chemisorption in fullerenes
and nanotubes has been considered earlier in Ref.
\onlinecite{curv}.

\section{Chemisorption of single hydrogen atom}

To model the hydrogen chemisorption we use a periodic supercell of
graphene containing 32 carbon atoms per each hydrogen atom,
similar to Ref. \onlinecite{Leht}. To consider hydrogen pairs, we
will use supercells with 50 carbon atoms for close pairs
(neighboring positions of hydrogen) and 72 carbon atoms,
otherwise. The density-functional theory calculations were
performed using the SIESTA code \cite{siesta1, siesta2} which was
successfully applied before to describe hydrogen on graphene
\cite{Yazyev1}. We used the same technical parameters of the
calculation as in Ref. \onlinecite{Yazyev1}.

\begin{figure}
\rotatebox{-90}{
\includegraphics[height=3.2 in]{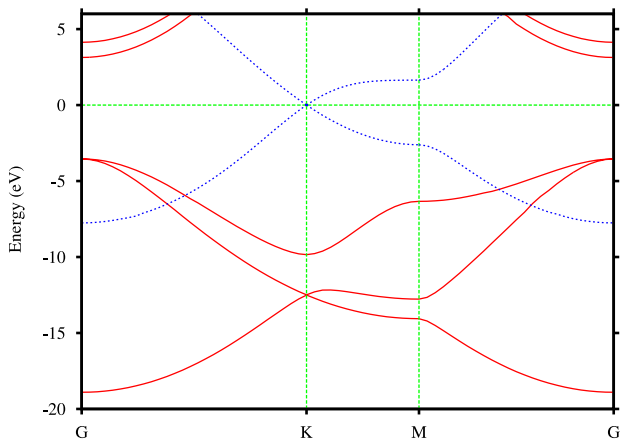}}
\caption{\label{fig1}(color online) Band structure of a single
graphene layer. Solid red lines are $\sigma$ bands
and dotted blue lines are $\pi$ bands.}
\end{figure}

\begin{figure}
\includegraphics[width=3.2 in]{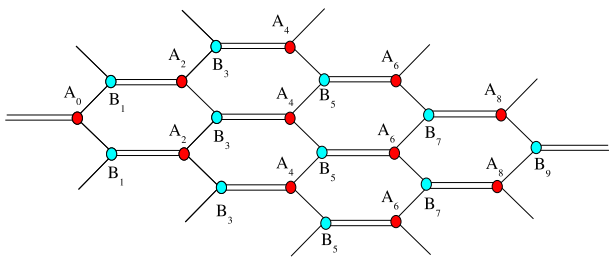}
\caption{\label{fig2}(color online) Crystallographic structure of
graphene. Red and blue circles show A and B sublattices,
respectively. Labels show the distance from A$_0$ carbon atom
(coordination sphere numbers).
All bonds in graphene are equivalent, the double bonds are marked
for convenience of comparison with other pictures.}
\end{figure}

To discuss chemisorption on graphene it is worth to remind its
basic electronic structure. Originally, carbon has two $2s$ and
two $2p$ electrons. These four electrons produce different kinds
of $sp$-hybridized orbitals \cite{pauling}. In graphene every
carbon atom is bounded with three other carbon atoms via $sp^2$
hybridization. There are three $\sigma$ orbitals placed in the
graphene plane with angle 120$^{\circ}$ and one $\pi$ orbital
along $Z$ axis in perpendicular direction. Figure \ref{fig1} shows
the band structure of pure graphene, with  three $\sigma$ bands
laying about 3 eV above and below the Fermi level, and $\pi$ band.
In diamond, all carbon atoms are connected via $sp^3$
hybridization with four $\sigma$ bands separated by a big gap.
Breaking $\pi$ bonds and producing additional $\sigma$ bond and,
thus, transition from $sp^2$ to $sp^3$ hybridization is the main
mechanism of chemisorption on graphene. The crystallographic
structure of graphene with two sublattices is shown in Fig.
\ref{fig2}. In pure graphene the sublattices are equivalent, but
if we bind one of carbon atoms (for example, A$_0$ in Figure
\ref{fig2}) with hydrogen we automatically break this equivalence.

\begin{figure}
\includegraphics[width=3.2 in]{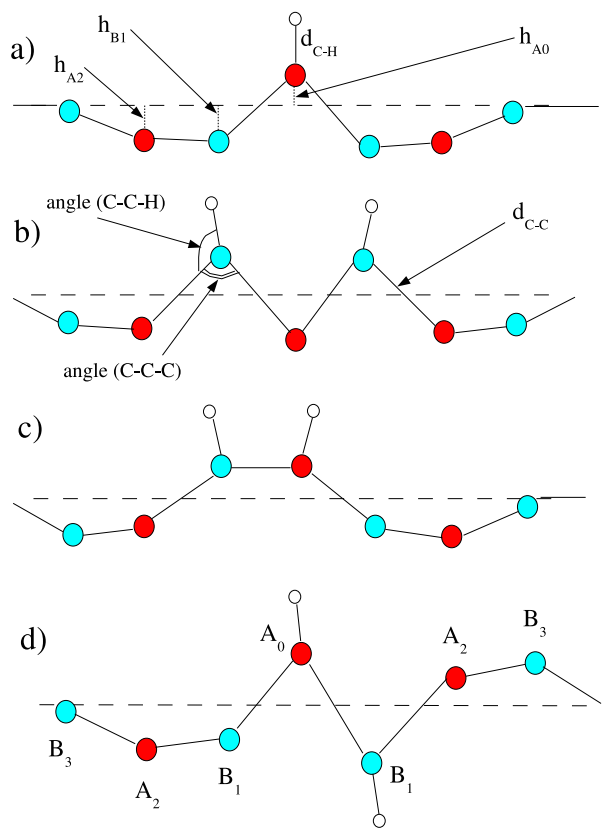}
\caption{\label{fig3}(color online) Picture of local distortions
of graphene at chemisorption of: (a) single hydrogen atom (A$_0$);
(b) two hydrogen atoms bonded with carbon atoms from the same
sublattice (A$_0$-A$_2$); (c) two hydrogen atoms bonded with
neighboring  carbon atoms from the same side of graphene sheet
(A$_0$-B$_1$); (d) two hydrogen atoms bonded with neighboring
carbon atoms from both sides of graphene sheet(A$_0$-B'$_1$). Red
and blue circles are carbon atoms from two sublattices, white
circles are hydrogen atoms.}
\end{figure}

To check the computational procedure, we reproduce first known
results \cite{Leht, Duplock, Yazyev1, Ito-MD} for single hydrogen
atom chemisorbed on graphene. In agreement with the previous
calculations we have found hydrogen-carbon distance about
1.1{\AA}, and shift of the carbon atom bonded with the hydrogen
one about 0.3{\AA} along $Z$ direction. One should stress,
additionally to the previous results, that the atomic distortions
are not negligible also for the second and third neighbors of the
hydrogen-bonded carbon atom A$_0$ (see Figure \ref{fig3}a).
Amplitude of the modulation of graphene sheet in the perpendicular
direction around the hydrogen atom was estimated as 0.4{\AA},
which is comparable with the height of intrinsic ripples on
graphene of order of 0.7{\AA} found in atomistic simulations
\cite{ripples}. The radius of the distorted region around hydrogen
atom turned out to be about 3.8{\AA}.

\begin{table*}
\caption{\\
Dependence of magnetic moments $M$ (in $\mu_B$), chemisorption
energies $E_{chem}$ (in eV), and geometrical parameters (see
Figure \ref{fig3}), in degrees and {\AA}, on configuration of
hydrogen (see Figure \ref{fig3}); $d$ are interatomic distances
and $h$ are heights of atoms from graphene plane.} \label{tab1}
\begin{ruledtabular}
\begin{tabular}{cccccccccc}
Configuration & $M$ & $E_{chem}$ & $h_{A0}$ & $h_{B1}$ & $h_{A2}$&
angle(C-C-H) & angle(C-C-C) & $d_{C-H}$ & $d_{C-C}$  \\ \hline
A$_0$ & 1.0 & 1.441 & 0.257 & -0.047 & -0.036 & 101.3 & 115.4 & 1.22 & 1.496   \\
A$_0$-A$_2$ & 2.0 & 1.406 & 0.285 & -0.040 & -0.096 & 102.7 & 116.6 & 1.132 & 1.483 \\
A$_0$-B$_1$ & 0.0 & 0.909 & 0.364 & -0.088 & -0.069 & 102.2 & 117.5 & 1.077 & 1.491 \\
A$_0$-B'$_1$ & 0.0 & 0.540 & 0.298 & -0.027 & -0.035 & 105.1 & 106.7 & 1.112 & 1.512 \\
\end{tabular}
\end{ruledtabular}
\end{table*}

Transformation of the $sp^2$ hybridization of carbon in ideal
graphene to the $sp^3$ hybridization in hydrogenated graphene
results in a change of the bond lengths and angles. A typical bond
length for $sp^2$ C-C bonds is 1.42{\AA} for graphene and graphite
and 1.47{\AA} for other compounds, and the standard bond angle is
120$^{\circ}$. For $sp^3$ hybridization, the standard value of C-C
bond length is 1.54{\AA}, and the angle is 109.5$^{\circ}$. A typical
value for the single C-H bond length is 1.086{\AA}. One can see in
Table \ref{tab1} that for single hydrogen atom the C-H bond length
is close to the standard value, but C-C-H and C-C-C angles are
intermediate between 90$^{\circ}$ and 109.5$^{\circ}$ and
120$^{\circ}$ and 109.5$^{\circ}$, respectively. Also, the length
of C-C bond is in between 1.42{\AA} and 1.54{\AA}. This means an
intermediate character of the hybridization between $sp^2$ and
$sp^3$.

A pictorial view of the reconstruction of chemical bonds, with the
breaking of double C=C bond and formation of single C-H bond, is
shown in Fig. \ref{fig4}.  For the case of single hydrogen atom
(Fig. \ref{fig4}a) this releases two unpaired electrons. One of
the electrons forms a new bond with hydrogen whereas the other is
unpaired. The latter is delocalized in some rather broad area on
lattice \cite{Yazyev1}. As a result, carbon becomes magnetic (see
the Table \ref{tab1}) and hydrogen atom also possess a small
magnetic moment about 0.12 $\mu_B$. In general, at the
chemisorption of single carbon atom the hybridization is still
rather close to $sp^2$. One has to consider another opportunities
which can lead to $sp^3$ bonding and possible gain in the
chemisorption energy.

\section{Hydrogen pairs on single-layer graphene}

\begin{figure}
\includegraphics[width=3.2 in]{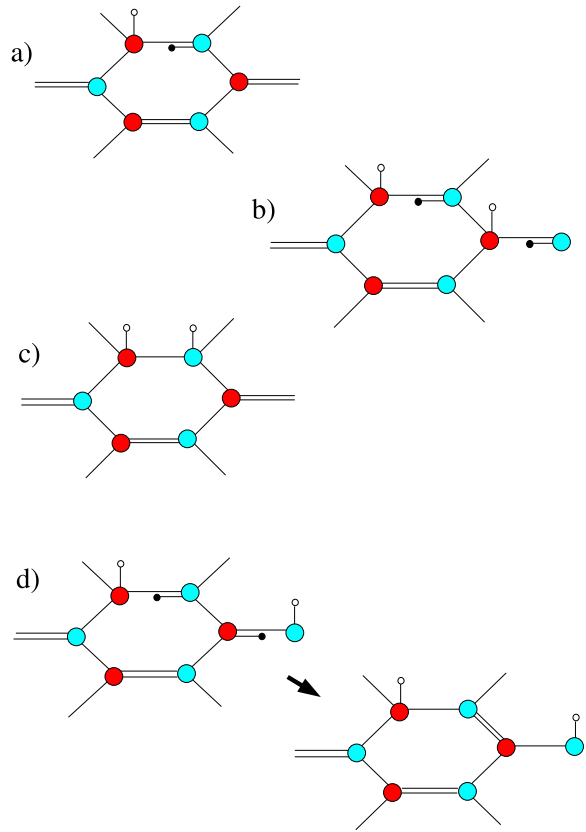}
\caption{\label{fig4}(color online) Sketch of chemical bonds for
chemisorption of hydrogen: (a) single hydrogen atom (A$_0$); (b)
two hydrogen atoms bonded with carbon atoms from the same
sublattice (A$_0$-A$_2$); (c) two hydrogen atoms bonded by nearest
carbon atoms (A$_0$-B$_1$); (d) two hydrogen atoms bonded by
next-nearest carbon atoms from different sublattices
(A$_0$-B$_3$). Big red (dark) and blue (light) circles
are carbon atoms from different sublattices,
small white circles are hydrogen atoms,
small black circles are unpaired electrons.}
\end{figure}

\begin{figure}
\rotatebox{-90}{
\includegraphics[height=3.2 in]{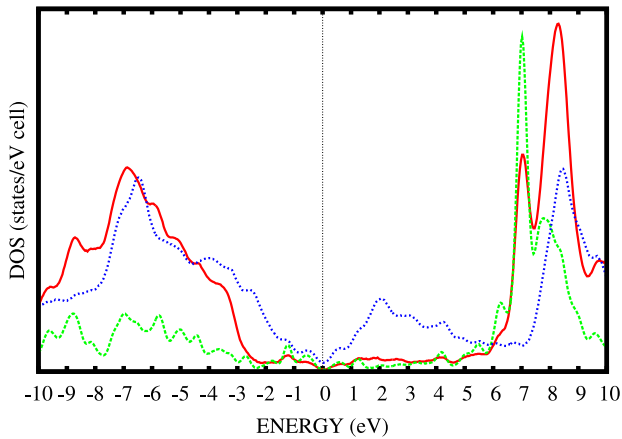}}
\caption{\label{fig5} (color online) Partial densities of states
for carbon atom bound with hydrogen (solid red line), hydrogen
atom (dashed green line), and distant carbon atom (from fourth
coordination sphere) (dotted blue line) for the case A$_0$-B$_1$'
configuration (see Fig. \ref{fig3}d). }
\end{figure}

There are four kinds of hydrogen pairs on graphene: hydrogen atoms
can be bonded by carbon atoms from the same sublattice of from
different sublattices, on one side from the graphene sheet, or
from both sides. We use the primed indices for the later case.
Computational results for chemisorption of hydrogen pairs are
presented in Fig. \ref{fig3} and in Table \ref{tab1}.
Chemisorption energy per hydrogen atom for the case A$_0$-A$_2$
(next-nearest-neighboring carbon atoms, both hydrogen atoms are
from the same side) is not significantly different from that for
single hydrogen, whereas chemisorption by carbon atoms from
different sublattices turns out to be much more energetically
favorable.

To understand the difference, one has to study what happens with
chemical bonds in all these cases. In Figure \ref{fig4}b, we can
see that for the case A$_0$-A$_2$ the situation is basically the
same as for the single hydrogen, namely, two broken bonds produce
two unpaired electrons with strong ferromagnetic coupling between
their spins (dependence of the exchange interactions from
interatomic distance was studied in detail in Ref.
\onlinecite{Yazyev1}). These electrons in the A$_0$-A$_n$ case are
not paired and produce new chemical bonds, the bond distances and
angles for A$_0$-A$_2$ being intermediate between those typical
for the $sp^2$ and $sp^3$ hybridization (see Table \ref{tab1}).

The situation A$_0$-B$_1$ is essentially different. One can see
from Fig. \ref{fig4}c that, when the double bond between A$_0$ and
B$_1$ carbon atoms transforms into the single one two unpaired
electrons appears and both of them participate in formation of
covalent bonds with the hydrogen atoms. For the case of more
distant carbon atoms, say, A$_0$ and B$_3$ we can see a similar
situation (Fig. \ref{fig4}d). Corresponding changes in the
electronic structure for this case is displayed in Fig.
\ref{fig5}. The density of states for carbon atoms bonded with
hydrogen in redistributed, decreasing in the region between -2.5
to 5 eV (the energy is counted from the Fermi level) and
increasing near $\pm{7}$ eV. These changes correspond to a
transition from $sp^2$ to $sp^3$ hybridization which makes
graphene-like electronic structure more ``diamond-like''
transforming the $\pi$ band crossing the Fermi level (see Fig.
\ref{fig1}) to fourth $\sigma$ band lying far from it. At the same
time, the electronic structure of fourth neighbors are very close
to electronic structure of pure graphene (Fig. \ref{fig5}). In the
case of chemisorption by carbon atoms from different sublattices
there are no unpaired electrons and no magnetism. In the work
\cite{Yazyev1} this situation was described as antiferromagnetic
which is not quite accurate as we believe. Actually, the local
magnetic moments just do not survive in this case. The absence of
unpaired electrons and broken bonds leads to chemisorption energy
gain in comparison with the A-A case described above.

Additionally, we can see in Table \ref{tab1} that the C-C bond
length for the case A$_0$-B$_1$ is close to the standard one for
$sp^3$ hybridization. However, the bond angles are closer to those
for $sp^2$ hybridization, and the chemisorption energy for the
case A$_0$-B$_1$ is higher than for A$_0$-B'$_1$. To understand
the difference, one has to investigate structural distortions of
graphene sheet. Chemisorption of hydrogen by A$_0$ carbon atom
induces its shift up perpendicular to the plane, together with
shifts of atoms B$_1$ and A$_2$ in the opposite direction. The
chemisorption on carbon B$_1$ atom shifts B$_1$ atom up and A$_0$
and A$_2$ atoms down. Therefore, for the case A$_0$-B$_1$ both
A$_0$ and B$_1$ carbon atoms move simultaneously in the same
direction. As a result, the bond angles become close to those
typical for $sp^3$ hybridization. On the contrary, in A$_0$-B'$_1$
case the chemisorption of hydrogen from the bottom by B$_1$ carbon
produces shifts up for A$_0$ and down for B$_1$ carbon atoms that
coincide with the lattice distortion for the bonding of hydrogen
by A$_0$ from the top. In the case A$_0$-B'$_1$ the lattice
distortions produced by chemisorption of each hydrogen atoms are
consistently working in the same direction providing the lowest
chemisorption energy and bond lengths and angles closest to the
standard ones for $sp^3$ hybridization (see Table \ref{tab1}).

The calculated dependence of the chemisorption energy on the
distance between carbon atoms bonded with hydrogen is presented in
Fig. \ref{fig6}. One can see that for all types of pairs the
chemisorption energy for the hydrogen atoms closer than 5{\AA} is
lower than for larger distances. Independently on the distance,
the nonmagnetic A-B pairs are more energetically favorable than
A-A pairs and than noninteracting hydrogen atoms. One can assume
therefore that observation of hydrogen-induced ferromagnetism
\cite{Yazyev1} is possible only for a very low concentration of
hydrogen when the distance between hydrogen atoms are higher than
12{\AA}. Our results  seem to be in a qualitatively agreement with
the experimental data on hydrogen chemisorption on highly-oriented
pyrolitic graphite (HOPG) \cite{pairs}. The pairs A$_0$-B$_3$ have
been observed which correspond to minimal energy for the one-side
hydrogenation of graphene, according to our results (see Fig.
{\ref{fig6}). Also, at hydrogenation of fullerenes C$_{60}$ the
pairs A$_0$-B$_1$ and A$_0$-B$_3$ (1,2 and 1,4, according to
chemical terminology) are usually observed (see, e.g., the review
\onlinecite{c60} and references therein). Instability of magnetic
state was observed experimentally for C$_{60}$H$_{24}$
\cite{C60H24}. Recent theoretical results for chemisorption on
single \cite{krash, SWCNTchem} and multiple-wall \cite{MWCNTchem}
carbon nanotubes are qualitatively similar to our results for
graphene.

\begin{figure}
\rotatebox{-90}
{\includegraphics[height=3.2 in]{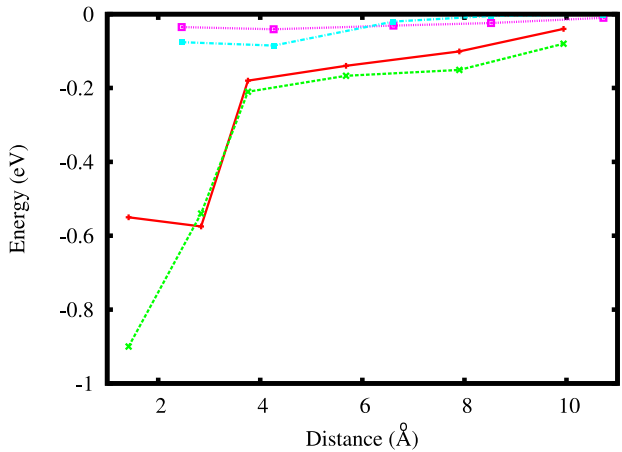}}
\caption{\label{fig6} (color online) Energy of hydrogen pair (per
atom) counted from the energy of single hydrogen atom as a
function of interatomic distance: A$_0$-B$_n$ -  solid red line with
crosses, A$_0$-B'$_n$ - dashed green line with crosses, A$_0$-A$_n$ -
dotted light blue line with filed squares,
A$_0$-A'$_n$ - dot-dashed violet line with empty squares.}
\end{figure}

\section{Hydrogen chemisorption on bilayer graphene}

Let us consider now hydrogen chemisorption on graphene bilayer. We
studied the chemisorption of single hydrogen atom and pairs of
hydrogen atoms placed on one and both sides of the bilayer. The
calculations have been performed for two different concentration
of hydrogen, that is, low (32 carbon atoms in each layer per
hydrogen atom) and high (8 carbon atoms in each layer per hydrogen
atom). Lattice distortions induced by the hydrogen turned out to
be different for the case of single-layer and bilayer graphene.
Whereas the shift of carbon atom bound with hydrogen is rather
similar in both cases, atomic displacements for the neighboring
carbon atoms are much smaller in the case of bilayer. This is not
surprising since interlayer coupling tends to make graphene more
flat, e.g., sheet corrugations are smaller for suspended bilayer
membrane than for the single-layer one \cite{meyer}.

Computational results are presented in Table \ref{tab2}. One can
see that for low hydrogen concentration the difference of
chemisorption energies  between single hydrogen atoms and the
pairs is smaller than for the case of single-layer. There are two
configurations which have a very close values of the energy for
low concentration of hydrogen, A$_0$-B$_1$ and A$_0$-B$_3$. For
the higher concentration, the latter configuration becomes
essentially more stable since the lattice distortions are more
homogeneous in this case. The effective interactions between
hydrogen atoms is more short-range in the case of bilayer and
already for the configuration A$_0$-B$_5$ the chemisorption energy
is almost equal to that of two single hydrogen atoms.

In the case of single-layer the hydrogen positions on different
sides of the graphene sheet are essentially more favorable than
those on the same side. Contrary, for the case of bilayer this
energy difference is small.

\begin{table*}
\caption{\\
Chemisorption energy $E_{chem}$ per hydrogen atom (in eV), height
$h$ of carbon atom bound with hydrogen up to the layer, and
interlayer distance $d$ (in {\AA}) for graphene bilayer for
different hydrogen concentrations and configurations of
chemisorbed hydrogen.} \label{tab2}
\begin{ruledtabular}
\begin{tabular}{ccccc}
Concentration & Configuration & $E_{chem}$ & $h$     & $d$     \\
\hline
Low           & A$_0$                & 1.28  & 0.639 & 3.237 \\
              & A$_0$-B$_1$ one side & 0.715 & 0.570 & 3.222 \\
              & A$_0$-B$_1$ both sides & 0.713 & 0.615 & 3.149 \\
              & A$_0$-B$_3$ one side & 0.720 & 0.477 & 3.237 \\
              & A$_0$-B$_3$ both sides & 0.733 & 0.453 & 3.237 \\ \hline
High          & A$_0$-B$_1$ one side & 0.885 & 0.445 & 3.174 \\
              & A$_0$-B$_1$ both sides & 0.850 & 0.426 & 3.041 \\
              & A$_0$-B$_3$ one side & 0.381 & 0.359 & 3.262 \\
              & A$_0$-B$_3$ both sides & 0.390 & 0.349 & 3.198 \\
\end{tabular}
\end{ruledtabular}
\end{table*}

\section{Hydrogen storage properties of graphene}

Chemisorption energy per hydrogen atom for the most favorable case
of A$_0$-B'$_1$ pairs presented in Table \ref{tab1} is not very
high. Another limiting case with much higher adsorption energy per
hydrogen atom corresponds to the case of fully hydrogenated
graphene which is close to a hypothetical compound graphane
\cite{graphan}. For the latter case, we have found bond lengths
1.526{\AA} for C-C bonds and 1.110{\AA} for C-H bonds, and bond
angles 102.8$^{\circ}$ and 107.5$^{\circ}$ for C-C-C and C-C-H
angles respectively, in a good agreement with the results of Ref.
\onlinecite{graphan}. The calculated values are close to the
standard ones for $sp^3$ hybridization, that is, 1.54{\AA} for the
length of C-C bonds and 109.5$^{\circ}$ for all angles. Value of
C-H bonds are also very close to the standard 1.09{\AA}.

We studied transition from single pairs to complete coverage
changing the supercell size. The dependence of the chemisorption
energy on the hydrogen concentration is shown in Fig. \ref{fig7}.
For fully hydrogenated graphene the mass percentage of hydrogen
(gravimetric energy density), is 7.8 which is over the target
value of DOE (United States Department of Energy) 6.5
\cite{Dillon1}. Another relevant characteristics for hydrogen
storage are energy barriers which are necessary to overcome to
start hydrogenation and dehydrogenation. They correspond to the
chemisorption energy per hydrogen atom for singe hydrogen pair and
for fully hydrogenated graphene, respectively. We found for these
quantities 0.53 eV (25.5 kJ/mol) and 0.42 eV (20.3 kJ/mol). The
latter value is close to the experimental one, 19.6 kJ/mol, for
hydrogenized nanotubes \cite{Dillon1}. These values look quite
reasonable in view of potential applications of graphene for the
hydrogen storage. Transformation of electronic structure with
increasing hydrogen concentration presented on the insets of Fig.
\ref{fig7}. Minimal mass hydrogen concentration which results in
opening of energy gap at the Fermi level is about 4.04 (50\%
coverage), the gap value being 1.75 eV. This seems to be,
potentially, an interesting prediction for experiment, although it
is not clear whether it is possible to stabilize this
configuration or not.

The computational results under discussion have been obtained in
the generalized gradient approximation, GGA, which is a common
practice for electronic structure calculations of H-C systems
\cite{Leht, Duplock, Yazyev1, ch4}. To estimate possible errors we
have calculated the desorption energy in the local density
approximation (LDA) as well. We have obtained the value 0.62 eV,
in comparison with the GGA result 0.42 eV so the difference is
essential. In more detail, the question was studied in Ref.
\onlinecite{ch4} with the conclusion that GGA is more reliable
than LDA for this kind of problems.

\begin{figure}
\rotatebox{-90}{
\includegraphics[height=3.2 in]{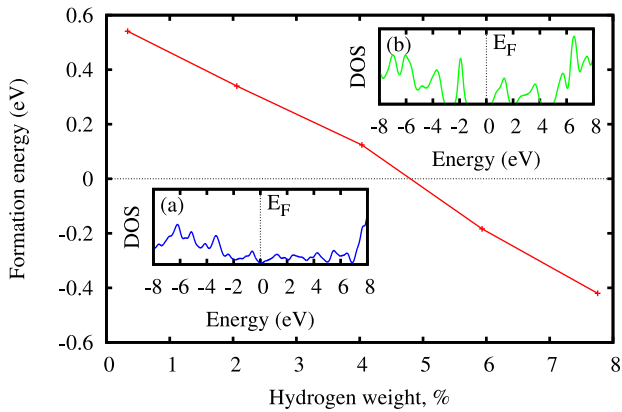}}
\caption{\label{fig7}(color online) Dependence of the
chemisorption energy per hydrogen atom on the mass hydrogen
concentration (gravimetric energy density). The insets shows
total densities of states for for (a) 2.06 and (b) 4.04 mass hydrogen
concentration.}
\end{figure}

\begin{figure}
\includegraphics[width=3.2 in]{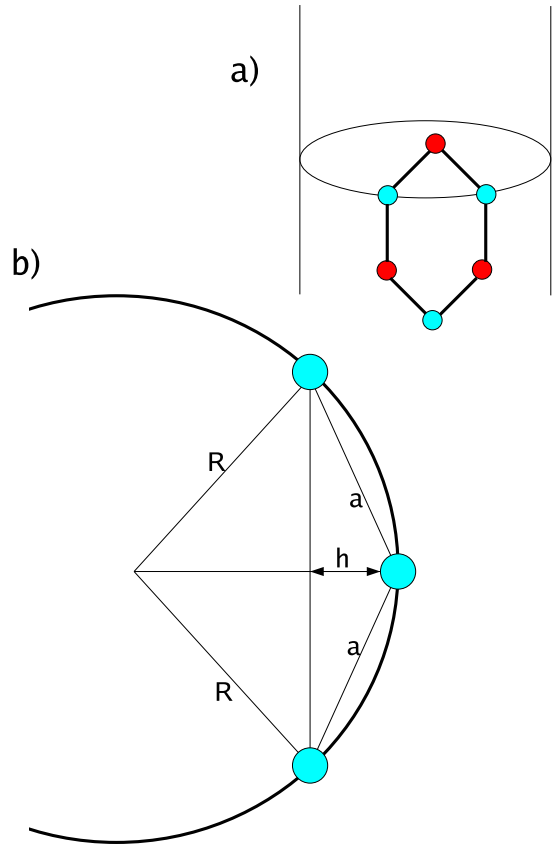}
\caption{\label{fig8}(color online) (a) Position of carbon
hexagons on surface of SWCNT. Red (dark) and blue (light) circles
are carbon atoms from different sublattices.
(b) Position of carbon atoms on radii of SWCNT.}
\end{figure}

In previous works \cite{Leht, Duplock} hydrogenation of graphene
was studied as a model of that of SWCNT. However, these two
situations are not identical due to curvature of the nanotubes. In
Fig. \ref{fig8} we sketch the SWCNT, value $h$ from Fig.
\ref{fig8}b corresponding to the sum of the values $h_{A0}$ and
$h_{A2}$ from Fig. \ref{fig3}a. In SWCNT $h = a^2/2R$, where $a$
is the lattice parameter for graphene, 2.46{\AA}, and $R$ is the
radius of nanotube. Typical diameters of the SWCNT 10$\div$15{\AA}
correspond to the values of $h$ from 0.605 to 0.375{\AA}. At the
same time, single hydrogen atom on graphene produces a distortion
with the value $h$ = 0.293{\AA}, that is lower than for the SWCNT
of standard diameter. This value is close to those for theoretical
estimations of maximum of the SWCNT diameter, 41.6
\cite{nanobreak1} and 49.9 \cite {nanobreak2}{\AA}.

On the other hand, multiple-wall carbon nanotubes (MWCNT) have
typical diameters about 50{\AA} and bilayer graphene can be a
reasonable model to study hydrogenation of the MWCNT. Moreover,
partial graphitization and presence of metallic catalysts strongly
influence on adsorption properties of SWCNT \cite{Dillon1} whereas
graphene is perfectly pure material. Another problems for
hydrogenation of nanotubes are how to provide an access of
hydrogen to their surface in an array \cite{array} and high enough
flip-into energy barrier \cite{bi2}.  Carbon 1s X-ray
photoemission spectra (XPS) of the SWCNT before hydrogenation,
after hydrogenation, and after dehydrogenation reported in Ref.
\onlinecite{1s} are all different that could be in part due to
defect formation whereas graphene has a very high vacancy
formation energy (up to 8 eV) which means much higher stability of
graphene under high temperatures and pressures.

\begin{table*}
\caption{\\
Dependence of chemisorption energy (in eV), interlayer distance
$d$, and geometrical parameters in {\AA} (see Fig. \ref{fig3}) on
numbers of graphene layers for 50\% hydrogenation of one side of
the top layer.} \label{tab3}
\begin{ruledtabular}
\begin{tabular}{ccccccc}
Number of layers & $E_{chem}$ & $d$ & $h_A$ & $h_B$ & $d_{C-H}$ &
$d_{C-C}$   \\ \hline
1 & 1.775 & -               & 0.106 & 0.143 & 1.158 & 1.475 \\
2 & 1.452 & 2.88            & 0.142 & 0.198 & 1.154 & 1.475 \\
5 & 1.621 & 3.124 and 3.353 & 0.133 & 0.116 & 1.164 & 1.468 \\
\end{tabular}
\end{ruledtabular}
\end{table*}

At last, we compared hydrogen storage properties of graphene and
graphite nanofibers (GNF), that is, very small graphite platelets,
with a size of order of 30$\div$500{\AA} \cite{GNF}. Raman spectra
for graphene multilayers become very close to ones for graphite
when number of layers is five or more so one can assume that
five-layered graphene is already similar to the bulk case
\cite{Ferarri}. To model the GNF we used therefore five-layer
graphene slab. Complete one-side hydrogenation of GNF, as well as
of graphene, is impossible and only 50\% hydrogenation of the top
layer is supposed to be the maximum (all carbon atoms from one of
the sublattices are bonded with hydrogens) that corresponds to
approximately 1\% of gravimetric energy density. The calculated
chemisorption energy per hydrogen for single-layer graphene with
the same gravimetric energy density is 0.32 eV lower than for
five-layer graphene. The maximum load of the five-layer graphene
is 2\% of the gravimetric energy density that is about four times
smaller than for the single-layer graphene. Results of calculation
for case of 50\% hydrogenated surface of graphene single-layer,
bilayer and graphite (five-layer of graphene) are presented in
Table \ref{tab3}. For all three structures chemisorption energies,
structural changes, magnetic properties and electronic structures
are essentially different. Some differences, e.g., in the length
of C-H bond are negligible, but many others (amplitude of bending
distortions) are significant. Detailed comparison of chemical and
structural properties of single-layer and multilayer  graphene
will be reported elsewhere.

\section{Conclusions}

We have performed the density functional calculations of electronic
structure, magnetic properties, and energetics of different
hydrogenated grephene layers. Our results support a suggestion
that graphene may be a promising material for the hydrogen
storage.  Equivalence of two sides of graphene distinguishes it
drastically from the nanotubes. We have shown that the most stable
configuration of low hydrogenated grephene layer corresponds to
the non-magnetic pair hydrogen atoms attached to the different A-B
sublattices of graphene from the differen sides. It is worth to
emphasize that single-layer \cite{vacan} or bilayer \cite{bilay,
Yazyev2} graphene should be rather carefully used as models of
structural and chemical properties of graphite and its derivates.
Also, a comparison of experimental results for graphite
\cite{magnogr} with the computational results for graphene sheets
\cite{Yazyev1} requires additional investigation.

\section{Acknowledgements}

We are grateful to A. Fasolino, M. van Setten and O.V. Yazyev for helpful
discussions. The work is financially supported by Stichting voor
Fundamenteel Onderzoek der Materie (FOM), the Netherlands.

\end{document}